# Preventing AI Deepfake Abuse: An Islamic Ethics Framework

1st Wisnu Uriawan
*Informatics Department*
*UIN Sunan Gunung Djati Bandung*
Jawa Barat, Indonesia
Corresponding author: wisnu_u@uinsgd.ac.id

2nd Imany Fauzy Rahman
*Informatics Department*
*UIN Sunan Gunung Djati Bandung*
Jawa Barat, Indonesia
imanyfauzyr123@gmail.com

3rd Muhamad Zidan
*Informatics Department*
*UIN Sunan Gunung Djati Bandung*
Jawa Barat, Indonesia
mzidanzidan045@gmail.com

4th Irma Rohmatillah
*Informatics Department*
*UIN Sunan Gunung Djati Bandung*
Jawa Barat, Indonesia
rohmatillahirma02@gmail.com

5th Muhammad Arkan Raihan
*Informatics Department*
*UIN Sunan Gunung Djati Bandung*
Jawa Barat, Indonesia
markanraihan@gmail.com

6th Irma Dwiyanti
*Informatics Department*
*UIN Sunan Gunung Djati Bandung*
Jawa Barat, Indonesia
irmadwiyanti.sky@gmail.com

*Abstract*—The rapid development of deepfake technology powered by AI has raised global concerns regarding the manipulation of information, the usurpation of digital identities, and the erosion of public trust in the authenticity of online content. These challenges extend beyond technical issues and involve complex moral dimensions, rendering conventional, technologically driven, and reactive management approaches insufficient to address underlying causes such as intent, ethical responsibility, and intangible social harm. In response to these challenges, this study aims to formulate a comprehensive Islamic ethical framework as a preventive approach to mitigate the misuse of deepfake technology. This study employed a Systematic Literature Review (SLR) guided by the Preferred Reporting Items for Systematic Reviews and Meta-Analyses (PRISMA), selecting ten primary sources published between 2018 and 2025 to identify ethical gaps, regulatory needs, and appropriate normative solutions. The analysis demonstrates that integrating the principles of *Maqāṣid al-Sharī'ah*, particularly *ḥifẓ al-'ird* and *ḥifẓ al-nafs*, provides a strong normative foundation for governing the responsible use of digital technology. Based on the findings, this study proposes three strategic recommendations: regulatory reforms that recognize the intangible and psychological harms resulting from reputational damage; strengthened technology governance grounded in moral accountability and the values of *'adl*, *amanah*, and transparency; and enhanced public digital literacy based on the principle of *tabayyun*. Overall, the findings suggest that the application of Islamic ethical principles shifts governance paradigms from punitive mechanisms toward preventive approaches that emphasize the protection of human dignity, the prevention of harm, and the promotion of the common good in the digital age.

## I. INTRODUCTION

In recent years, rapid advances in Artificial Intelligence (AI) have opened a new chapter in human-machine interaction. AI is no longer just an automation tool; it has become integral to how we work, communicate, and perceive digital reality. However, these advances also raise complex ethical dilemmas, particularly as AI's ability to generate highly similar synthetic content poses serious risks of misuse. As generative models such as GPT and diffusion models become more sophisticated, the line between reality and simulation is increasingly blurred. AI is now capable of imitating facial expressions, voices, and even human communication patterns with high precision. This phenomenon raises concerns about potential moral and social misconduct, including information manipulation, identity theft, and the spread of visual-based misinformation. [1]. One of the most problematic manifestations of this phenomenon is deepfakes, a technique based on Generative Adversarial Networks (GANs) that can create or modify audio-visual content to appear authentic. In legitimate contexts, deepfakes can provide positive benefits, such as in film production, educational simulations, or digital history restoration. However, when misused, this technology can compromise privacy, defame reputations, and even erode public trust in digital media. [2], [3].

Despite these severe risks, current reactive and technologically driven management methods—such as detection algorithms or quick regulatory responses often fail to address the core moral issues underlying deepfake abuse, including the user's intent, the ethical vacuum in development, and the long-term, intangible social impact. Thus, the challenge is not only technical but fundamentally ethical. This paper argues that an effective, preventive solution requires a comprehensive ethical framework that focuses on human morality and responsibility, moving beyond mere detection. Based on this premise, we propose an exploration of Islamic ethics, with its established principles of truth, justice, and responsibility, as a robust foundation to construct a human-centered, value-based governance model for AI and deepfake technology.

Globally, the misuse of deepfakes has raised serious concerns due to its role in spreading disinformation and manipulating public opinion. Deepfake content is used for political gain, propaganda, and reputational attacks against individuals or institutions. [4]. As a result, the line between fact and fabrication is becoming increasingly blurred, weakening the principle of trust that underpins social interaction in the digital space. This problem is further complicated by the lack of comprehensive ethical and legal standards for addressing the misuse of deepfakes. Several countries have developed AI governance frameworks that emphasize the moral responsibility of technology developers and providers, but their effectiveness depends heavily on the integration of law, ethical values, and public awareness. [5].

Given the limitations of purely technical and regulatory solutions, the need for a foundational ethical approach becomes urgent. This study posits that the Islamic ethical system offers a robust, preventative framework due to its emphasis on core moral principles that are highly relevant to the digital age, such as *ṣidq* (truthfulness) and *amanah* (trust/responsibility). By integrating these ethical values into the governance of deepfake technology, especially through the lens of *Maqāṣid al-Sharī'ah* (the higher objectives of Islamic law), we can shift the focus from merely detecting harm to proactively protecting fundamental human interests, such as honor, life, and intellect, thereby ensuring AI serves the common good and prevents moral misconduct.

Beyond legal and social aspects, the spread of deepfakes also has psychological implications. Studies show that repeated exposure to synthetic content can reduce people's trust in visual evidence, increase skepticism of online information, and create mass confusion [6]. This situation demonstrates that addressing deepfake abuse through repressive legal regulations alone is not sufficient. A more fundamental ethical approach is needed one that emphasizes moral responsibility, honesty, and respect for human dignity in the development and distribution of AI-based technology [7]. In this context, Islamic ethics offers a significant contribution through moral principles that emphasize benefit *(maṣlaḥah)*, justice *('adl)*, *(amanah)*, and protection of honor and privacy *(ḥurmah)*. These values can serve as a foundation for formulating more humane and equitable AI ethics policies and governance [8].

In Indonesia, the misuse of deepfakes, including voice impersonations of public figures, non-consensual pornography, online fraud, and disinformation, has become a real threat. This situation creates epistemic uncertainty, where people are no longer sure whether the digital content they see or hear is an authentic representation or a fabricated manipulation [9], [10]. Therefore, developing an ethical AI governance framework that integrates an Islamic ethical perspective is imperative. In a global context, Islamic values offer an alternative to the dominant Western AI ethics paradigm, emphasizing information honesty, identity protection, and the prohibition of harming others [11].To this end, this study specifically adopts an expanded framework of *maqāṣid al-sharī'ah*. In addition to the five classical principles *(al-kulliyāt al-khamsah)*, this study emphasizes protection of honor *(ḥifz al-'ird)* as a sixth, independent and essential pillar in today's digital context. These principles, especially *ḥifz al-'ird* as well as protecting reason *(ḥifz al-'aql)* and protecting offspring *(ḥifz al-nasl)*, can be used as a fundamental preventive framework to avoid the misuse of technology that has the potential to harm the public interest.

Studies such as AI in Islamic Ethics: Towards Pluralist Ethical Benchmarking for AI" and "Analysis of Islamic Law and Ethics on the Use of Deepfake Technology by Teenagers" demonstrate how Islamic principles can enrich the global dialogue on AI ethics and serve as a reference in addressing the misuse of technology, particularly among the younger generation [12], [13]. Therefore, this paper aims to develop an Islamic ethical framework as a preventive tool against the misuse of AI-based deepfakes. This approach emphasizes that advanced technology cannot be separated from moral and spiritual responsibilities. Integrating Islamic values into AI governance is expected to establish a new paradigm that guides technological innovation so that it remains aligned with the public interest, justice, and humanity.

Conceptually, the urgency of this research lies in the need to fill the gap between positive legal regulations and normative ethical foundations derived from Islamic teachings. To date, most studies on deepfake abuse have focused on technical aspects and secular policies, while the spiritual and religious moral dimensions have not been comprehensively integrated. Therefore, this research offers an alternative approach by examining how Islamic ethical values can serve as a preventive measure in building digital moral awareness, strengthening the social responsibility of AI developers, and encouraging the development of policies that support the protection of human dignity in the era of AI.

## II. Related Work

Studies on the relationship between AI and Islamic values are growing as technology integrates more and more into human life. Numerous studies have explored the ethical, legal, and *maqāṣid al-sharī'ah* aspects of AI applications, with the goal of developing systems that are not only technically intelligent but also align with Islamic moral and spiritual principles. A literature review indicates that one of the most pressing issues is how to ensure moral and legal accountability for increasingly autonomous AI entities. This challenge potentially conflicts with the traditional concept of the subject of law and ethics in Islam, which places moral responsibility on humans as vicegerents on earth. To address this gap, several studies have addressed how Islamic legal and ethical principles can be adapted to regulate the development of AI that mimics human behavior [14]. Other research emphasizes the need for a pluralistic ethical approach to AI development by incorporating Islamic principles, through an ethical assessment framework that places the values of *('adl)*, responsibility, and benefit *(maṣlaḥah)* as benchmarks in AI design and implementation [12]. This approach broadens the global discourse on the ethical governance of AI with a cross-cultural and cross-religious perspective.

Studies on the contribution of *maqāṣid al-sharī'ah* show that AI can be an instrument of benefit if used in accordance with sharia values such as *('adl)* and the common good *(maṣlaḥah 'āmmah)* [15]. This concept is relevant in the modern context, where innovation often gives rise to moral dilemmas such as data misuse and digital information manipulation. Other research highlights AI ethics through the lens of the *Sunnah* of the Prophet Muhammad (peace be upon him), which emphasizes the value of wisdom *(ḥikmah)* and responsibility in the use of technology. This principle encourages the application of Islamic values in the design of algorithms that directly impact human life, so that technology is assessed not only in terms of efficiency, but also in terms of the human values and spirituality it contains [16]. This approach emphasizes that technological progress should not be separated from the moral foundation and prophetic values that guide humans towards a balance between this world and the hereafter. Thus, the development of ethical AI according to Islam becomes a means to realize a digital civilization that is just and oriented towards the welfare of the people.

Several recent studies confirm that the integration of Islamic ethics in AI development can strengthen the moral foundation of global technology governance. This approach positions humans not only as users but also as guardians of digital morality responsible for the social impact of each innovation. By adopting the principles of justice, trustworthiness, and honesty in the design of AI systems, Muslim communities can play an active role in shaping an inclusive and welfare-oriented technology ecosystem. This demonstrates that the application of *maqāṣid al-sharī'ah* in AI is not merely a normative effort, but rather a concrete strategy to ensure that technological development remains aligned with humanitarian values, spirituality, and social balance in an increasingly complex digital era.

In the digital economy, the application of AI has also been studied as an instrument to ensure compliance with Sharia principles in Islamic financial institutions. This technology has been proven to increase transparency and accountability in Islamic accounting practices and strengthen public trust in the Islamic financial system [17]. This demonstrates that AI can be directed to support Islamic moral principles, rather than contradict them. An Islamic virtue based approach to AI development has also been proposed through the integration of values such as honesty, responsibility, and fairness into algorithms, so that AI ethics is understood not only as compliance with the law but also as an effort to shape the system's ethical and moral character [18]. Other research emphasizes the importance of Islamic digital ethics in the development of AI technology, by positioning the *maqāṣid al-sharī'ah* as a moral guide for developers to create digital systems that are fair, safe, and respectful of human dignity *(ḥurmah al-insān)* [19]. This strengthens Islam's position as a relevant source of values for the development of global ethics in information technology.

Recent studies also show that the development of digital technology has significant potential to shape the religious behavior and perceptions of modern society. The use of artificial intelligence in the context of Islamic da'wah and education is beginning to be seen as a strategic opportunity to expand the reach and effectiveness of conveying moral messages and Islamic values. Furthermore, a study of Islamic content on social media highlights how digital technology can be utilized positively to spread Islamic values. Through the use of web scraping methods and Pearson correlation analysis of da'wah videos on Instagram, the study demonstrates that the use of technology can strengthen digital da'wah and Islamic literacy, provided it is used ethically and responsibly [20]. The relevance of these findings lies in the importance of Islamic digital ethics in ensuring that AI-based innovations, including generative technologies like deepfakes, are not used for manipulation or moral deviation.

Overall, these various studies demonstrate the significant potential of integrating AI and Islamic values to shape a more ethical and equitable technological paradigm. The collaboration between Sharia principles and Information and Communication Technology (ICT) innovation can produce artificial intelligence systems that are not only technically superior but also oriented towards the welfare of the community. In the context of technological misuse such as deepfakes, an understanding of Islamic ethics becomes increasingly important in building a preventive framework that rejects manipulation, disinformation, and violations of individual honor and privacy.

III. METHODOLOGY

This study uses a literature review method to analyze how Islamic ethical values can be applied to address the misuse of AI based deepfake technology. This method was chosen because it allows researchers to collect, review, and synthesize relevant previous research findings without conducting direct experiments. This approach focuses on exploring and analyzing credible scientific sources to gain an in-depth understanding of the development of Islamic ethical concepts in the context of AI and deepfake technology. Through the literature review, researchers identify various perspectives, theories, and moral principles proposed by experts and then compare them to identify patterns, gaps, and relevant research directions. Thus, this method not only provides a strong theoretical foundation but also builds a systematic and contextual ethical framework to address the ethical challenges posed by advances in modern digital technology [21]. The stages in this method are explained in Figure 1, as follows:

*A. Identifying Research Topics and Objectives*

The primary focus of this research is to examine in depth how an Islamic ethical framework can function as a preventive instrument to prevent and minimize the misuse of AI based deepfake technology. The goal is not only to diagnose risks such as the spread of disinformation, fraud, and privacy violations, but also to offer normative solutions rooted in religious values. Through this approach, the research seeks to demonstrate that Islamic principles such as honesty *(ṣidq)*, *(amanah)*, and benefit *(maṣlaḥah)* can be applied contextually to address ethical challenges in the digital age. Thus, an Islamic ethical framework serves not only as a moral guideline but also as a practical guide for formulating policies, improving digital literacy, and strengthening social awareness to create a safe, just, and humanitarian-based technology ecosystem.

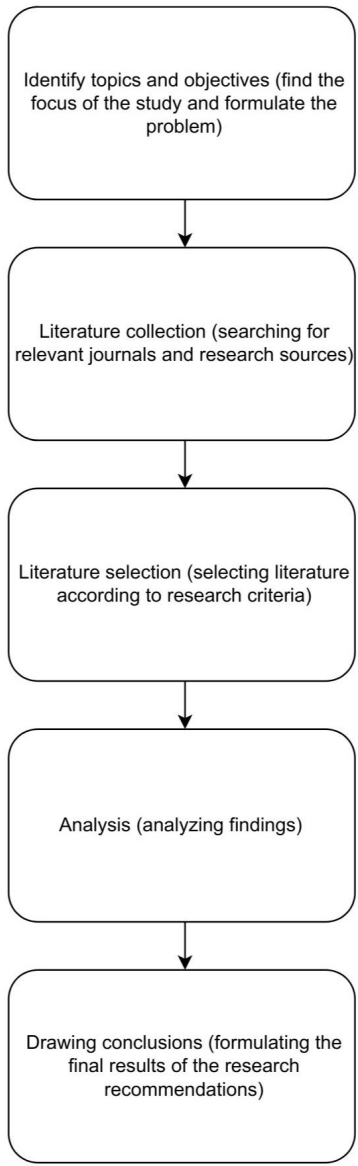

Fig. 1. Literature Review Method

This research specifically aims to formulate an integrated Islamic ethical model capable of providing moral and legal boundaries for preventing the creation and dissemination of damaging deepfake content. This framework ensures that the use of AI remains aligned with universal Islamic principles, such as preserving human dignity *(ḥurmah)*, ensuring *('adl)*, and prioritizing the public good *(maṣlaḥah)*. Thus, the resulting framework serves as both an ethical guideline and a theoretical foundation for developing technology policies and regulations with an Islamic perspective [22].

*B. Literature Collection*

The literature collection process was conducted systematically, utilizing reputable international scientific databases such as Springer, IEEE, Elsevier, and Taylor & Francis. This source selection was crucial to ensure the validity and depth of information related to the technical, governance, and ethical aspects of AI and deepfake technology from a global and current scientific perspective [23]. Furthermore, the selection process was conducted using relevant keywords such as "Islamic ethics", "AI governance", "deepfake prevention", and "ethical framework" to ensure a comprehensive scope of the study. Each selected literature was then evaluated based on its relevance, methodology, and contribution to the development of an Islamic ethical framework in the context of digital technology. This approach ensures that the literature review is not only comprehensive from a technological perspective but also rich from an Islamic philosophical and theological perspective. Each selected literature was then evaluated based on its relevance, methodology, and contribution to the development of an Islamic ethical framework in the context of digital technology. Thus, this study builds a strong theoretical foundation, balancing the discourse of Islamic ethics with modern technological ethics, and is able to offer a contextual and applicable perspective in addressing the challenges of AI ethics in the contemporary era.

*C. Selection of Relevant Literature*

The literature selection process was conducted rigorously and systematically to ensure that only publications with a high level of relevance, credibility, and academic contribution were analyzed in this study. The primary objective of this stage was to eliminate sources that were speculative or did not support the focus of Islamic ethics in the context of the misuse of AI based deepfake technology. The selection process involved two main stages: initial screening and in-depth eligibility assessment, guided by the PRISMA method guidelines to ensure a transparent and replicable selection process.

In the initial screening stage, researchers reviewed the title and abstract of each publication to assess its initial relevance to the research topic. Sources that did not explicitly address AI, ethics, or Islam were excluded from the list. Then, an in-depth assessment of the full text was conducted to assess methodology, analytical depth, and contribution to the understanding of Islamic ethics and digital technology. Each publication was also checked for duplication, academic validity, and publisher reputation. The literature selection criteria included:

1) Discussion Focus: Literature should explicitly address the issue of AI or deepfakes in relation to ethics. This criterion includes studies examining moral responsibility, algorithmic fairness, digital privacy, and the social impacts of AI misuse. Research that addresses only technical aspects without addressing ethical or moral aspects is excluded. This is crucial to ensure that the selected sources are truly relevant to the analysis of Islamic values in the context of AI.

2) **Islamic Values Content**: Selected publications must contain a discussion of Islamic ethical values related to modern technological issues. The values studied include trust and responsibility for information, public benefit *(maṣlaḥah)*, social and algorithmic *('adl)*, and protection of human dignity and privacy *(ḥurmah)*. Literature explaining the application of these values in the context of AI governance, digital sharia law, or technology ethics is a top priority. This criterion ensures that research analysis has a strong normative basis in the principles of *maqāṣid al-sharī'ah*.

3) **Publication Period**: The timeframe of 2018–2025 was chosen to ensure the relevance of the discussion, given the significant developments in generative AI technology over the past seven years. Research from this period reflects the current dynamics in the global discourse on AI ethics and the misuse of deepfakes. The selection of literature from this period also allows for an analysis of Islamic intellectual responses to cutting-edge technological phenomena such as GANs, Large Language Models (LLMs), and multimodal AI systems. [24].

Furthermore, all publications meeting the above criteria were verified through publisher reputation (e.g., Springer, IEEE, Elsevier, Taylor & Francis) and journal indexing status (Scopus or Sinta). This process ensured that the literature used had high academic validity and was representative of the development of Islamic ethics discourse in the context of AI. The final selection results yielded a comprehensive and relevant body of literature to support the analysis and synthesis stages of this research.

### D. Literature Analysis and Synthesis

This stage involved an in-depth content analysis of the selected literature to identify key patterns, commonalities, and differences in approaches emerging from various previous studies. This analysis was conducted by critically examining how researchers view the deepfake phenomenon from ethical, legal, and social perspectives, and how they propose mitigating measures to address its negative impacts. Some literature highlighted the dangers of disinformation and the decline of public trust as primary threats, while others emphasized the importance of regulation, digital literacy, and moral awareness as long-term solutions. By combining these diverse perspectives, the study sought to find common ground between technological approaches and normative values that could strengthen ethical oversight of the use of AI. This analysis provided a crucial foundation for formulating the proposed Islamic ethical framework.

The synthesis of the analysis results emphasizes the integration of the principles of *maqāṣid al-sharī'ah*, namely the protection of religion *(ḥifz al-dīn)*, life *(ḥifz al-nafs)*, reason *(ḥifz al-'aql)*, *(ḥifz al-nasl)*, and property *(ḥifz al-māl)* as preventive pillars against the misuse of AI. In addition, moral values such as honesty and *(amanah)*, anti-manipulation*(ṣidq)*, and *('adl)* serve as operational guidelines for both developers and users of technology [25]. The integration of these principles is expected to build an ethical paradigm that not only protects society from the negative impacts of deepfakes but also encourages the development of AI technology that is fair, transparent, and oriented towards the public good.

### E. Synthesis and Derivation of Ethical Recommendations

This stage represents the final synthesis phase of the systematic literature review, in which the selected studies are integrated to derive a coherent ethical framework grounded in the principles of *maqāṣid al-sharī'ah* and universal moral values. Rather than serving as a concluding statement, this synthesis functions as an analytical step to translate recurring ethical patterns identified in the literature into structured normative recommendations. Based on this synthesis process, three key ethical and policy-oriented recommendation themes emerge from the reviewed literature:

1) **Regulatory dimension**: The literature highlights the need for amendments to the Information and Electronic Transactions Law to recognize reputational harm and immaterial damage, including mechanisms for honor protection and reputation restoration.

2) **Technological governance dimension**: Several studies emphasize the importance of ethical audits for generative AI models to ensure alignment with the principle of *'adl*, particularly in mitigating algorithmic bias.

3) **Educational dimension**: The reviewed literature consistently recommends the development of digital literacy curricula grounded in the values of *tabayyun* and *amanah*, in collaboration with educational and religious institutions.

Collectively, these themes illustrate how the selected literature converges toward a preventive ethical orientation that integrates regulatory, technological, and educational dimensions within an Islamic moral framework [26].

To illustrate the systematic process of identifying, selecting, and synthesizing the literature used in this study, the PRISMA guidelines were employed as a visual reference, as shown in Figure 2. The PRISMA Diagram illustrating the systematic literature review process. presents the number of records identified, screened, assessed for eligibility, and ultimately included in the analysis.

1) **Identification and Screening Phase**: The literature review began with an identification process that yielded 44 records in total, consisting of 26 records retrieved from academic databases and 18 additional records identified through websites, organizational reports, and citation tracking. During the initial screening, 14 records were removed due to duplication (8 records) and other exclusion reasons (6 records). The remaining 30 records were then screened based on title and abstract relevance, resulting in the exclusion of 12 records and leaving 18 reports for further assessment.

2) **Eligibility Assessment Phase**: Of the 18 screened reports, 15 full-text articles were successfully retrieved,

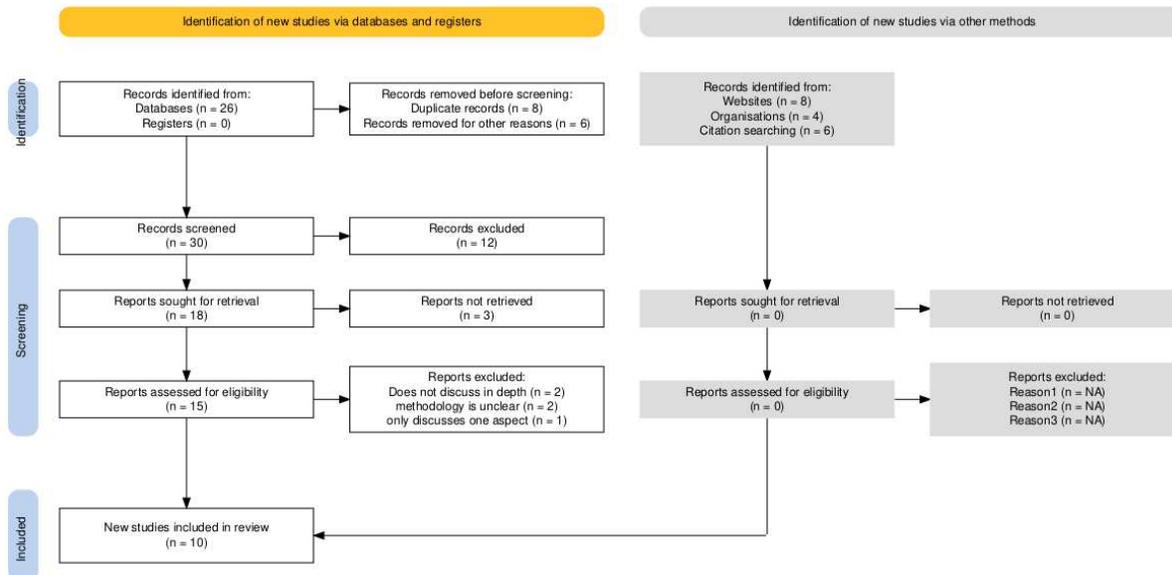

Fig. 2. PRISMA Diagram

while 3 were inaccessible. These full-text articles were further evaluated based on predefined relevance and quality criteria. As a result, 5 reports were excluded: 2 did not sufficiently address the research focus, 1 lacked a clear methodological approach, and 2 examined only a single dimension of the issue.

3) Included Studies: Following the completion of all PRISMA stages, 10 studies were identified as meeting the inclusion criteria. These studies constituted the final corpus used for qualitative content analysis and ethical synthesis in this research, ensuring that the analytical framework was grounded in relevant and credible scholarly sources.

## IV. Result and Discussion

Based on the PRISMA-guided systematic literature review, a total of 10 core studies were identified as the most relevant and credible sources for analyzing the ethical challenges of deepfake technology from an Islamic perspective. These studies were published between 2024 and 2025, indicating that the topic remains highly contemporary and actively discussed in recent academic discourse.

As summarized in Table I, the selected literature spans multiple disciplinary perspectives, including Islamic law, digital ethics, artificial intelligence governance, and social impacts of deepfakes. Several studies (e.g., Ramadhani, Afif, and Rohmawati) focus on legal protection and regulatory responses to deepfake abuse, particularly in the Indonesian context. Other works (such as Judijanto and Basri) examine ethical AI governance frameworks, emphasizing the relevance of Islamic values and *Maqasid al-Shari'ah* in regulating emerging technologies.

In addition, a number of studies address the societal and ethical dimensions of deepfake technology. Research by Boediman highlights the erosion of public trust and media credibility caused by deepfake manipulation, while works by Hadiyanto and Akbar emphasize ethical education and moral awareness, especially among youth and within educational institutions. Studies by El-Hady and Firdaus further explore broader Islamic ethical perspectives on artificial intelligence, reflecting diverse viewpoints within Muslim communities.

Overall, the results indicate that existing literature consistently recognizes deepfakes as a multidimensional problem, involving legal gaps, ethical concerns, and social risks. The selected studies collectively demonstrate a growing scholarly consensus on the need for value-based and preventive approaches, providing a solid empirical foundation for further interpretative discussion and the development of an Islamic ethical framework in the subsequent section.

### A. Literature Review Synthesis Results

An analysis of ten carefully selected core literatures demonstrates a consensus among researchers regarding the limitations of technological approaches and ethical frameworks developed in the West. While emphasizing justice, transparency, and accountability, these approaches are deemed incapable of addressing the root moral issues underlying the emergence of the deepfake phenomenon. Their orientation remains reactive and focuses on technical regulations, rather than on fostering preventative individual ethical awareness.

Further study indicates that the *Maqasid al-Shari'ah*, or primary objectives of Islamic law, have the potential to serve as a strong normative foundation for developing just and morally sound artificial intelligence governance. This framework provides a more comprehensive value orientation because it integrates the dimensions of benefit, justice, and protection of human dignity within a holistic ethical system.

TABLE I
STATE-OF-THE-ART LITERATURE REVIEW ON AI-BASED DEEPFAKES AND ISLAMIC ETHICS

| No | Author | Year | Title |
|----|--------|------|-------|
| 1 | Ramadhani, E.R. | 2025 | Integrating Islamic Values with the Right to Be Forgotten: A Legal Approach to Addressing Deepfake Pornography in Indonesia |
| 2 | Hadiyanto, R. | 2025 | Analisis Hukum Islam dan Etika Terhadap Penggunaan Teknologi Deepfake oleh Remaja |
| 3 | Afif, E. | 2025 | Perlindungan Hukum Terhadap Korban Penyalahgunaan Deepfake dalam Perspektif Islam |
| 4 | Judijanto, L. | 2025 | Implementation of Ethical Artificial Intelligence Law to Prevent Deepfakes |
| 5 | Boediman, E.P. | 2024 | Exploring the Impact of Deepfake Technology on Public Trust and Media Manipulation: A Scoping Review |
| 6 | El-Hady, E.H.F. | 2024 | Pandangan Islam terhadap Etika Kecerdasan Buatan (Artificial Intelligence) dalam Kehidupan Sehari-hari |
| 7 | Basri, W. | 2025 | Transforming Ethical Regulation of Artificial Intelligence in Islamic Banking: A Maqashid Shariah Perspective in the Digital Era |
| 8 | Akbar, A. | 2025 | Etika Islam terhadap Kecerdasan Buatan: Kajian Maqashid Syariah dalam Implementasi AI di Lembaga Pendidikan Islam |
| 9 | Rohmawati, I. | 2024 | Urgensi Regulasi Penyalahgunaan Deepfake sebagai Perlindungan Hukum |
| 10 | Firdaus, M.A. | 2025 | Islam and Artificial Intelligence: Perspectives from Traditionalist and Modernist Muslim Communities in Indonesia |

Specifically, the deepfake crisis is directly addressed by two primary components of the *Maqāsid al-Sharī'ah*: *ḥifz al-'ird* and *ḥifz al-nafs*. The creation and distribution of deepfakes for defamation, blackmail, or political manipulation constitutes a clear violation of *ḥifz al-'ird* by inflicting immediate and often permanent damage to an individual's reputation, which is considered a sacred asset in Islam. This ethical violation is further compounded as the resultant psychological trauma, social exclusion, and economic damage triggered by the deepfake content directly threaten the holistic well-being of the individual, thus encroaching upon *hifz al-nafs*. By explicitly classifying such intangible damages—like reputational harm and psychological distress—as a moral transgression that must be prevented, the Islamic framework transcends the limitations of conventional law, which often struggles to quantify and penalize non-material digital harm. This approach necessitates a shift towards preventative ethical implementation, focusing on cultivating *amanah* among developers and promoting *tabayyun* among users, ensuring that technology serves the higher objective of preserving human welfare (*maṣlaḥah*) above all else.

Furthermore, applicable Islamic moral values such as *(amanah)*, prohibition of lying *(kizb)*, and verification of information verification of information grounded in *tabayyun* emerge as ethical principles with high relevance in the digital context. These values not only serve as normative guidelines, but can also be practically applied in building a digital culture that is ethical, transparent, and has integrity.

*B. Legal and Regulatory Approach Based on Maqasid al-Shari'ah*

A thorough review of various articles indicates that the legal framework for handling digital crimes such as deepfakes must shift from a legalistic-positivistic approach to a more comprehensive model based on human welfare values. An approach based on *Maqasid al-Shari'ah* is considered capable of providing an ethical and normative basis for developing digital regulations because it offers the principle of ḥifz al-'ird, which has not been sufficiently addressed in positive law.

The limitations of the legalistic-positivistic approach stem from its tendency to focus on material, measurable damages and retroactive punishment, leaving a significant gap in addressing the intangible, psychological, and systemic harms caused by deepfakes, such as loss of social trust and trauma. In contrast, the integration of *ḥifz al-'irḍ* within the regulatory framework provides a powerful, pre-emptive moral shield against deepfake abuse. This principle mandates that any act, digital or otherwise, that leads to defamation, humiliation, or reputational damage is a severe moral transgression that must be prevented at the source, moving the focus from the act of detection to the ethical intent of the technology creator and user. Consequently, regulating deepfakes through Maqāṣid al-Sharī'ah promotes a culture of *amanah* in the digital ecosystem, demanding that all stakeholders from AI developers to end-users uphold moral obligations that prioritize the preservation of human honor and prevent the fabrication of falsehoods (*kidhb*), thereby safeguarding the *maslahah* in the digital public sphere.

In the case of deepfake abuse, the principle of *hifz al-'ird* holds crucial relevance because the impacts are not only material but also immaterial, such as defamation, mental trauma, social pressure, and long-term stigma for victims. These efforts have proven difficult to achieve through traditional legal systems, which often focus solely on physical or economic losses. This is where the values of *Maqasid* can enrich the legal framework by providing moral and spiritual legitimacy that protecting digital honor is a fundamental right of every individual.

The concept of the "right to be forgotten", which is gaining acceptance in international digital law discussions, is highly suitable for inclusion within the Islamic legal framework in Indonesia. This principle aligns with the Maqasid because it provides a restorative means for victims to erase digital traces that damage their identities. Incorporating this concept will encourage the legal system to not only prosecute perpetrators but also ensure the restoration of victims' dignity through digital rehabilitation and psychosocial support. Furthermore, the need for policy reform is increasingly apparent given the transnational, rapid, and complex nature of digital crime. Traditional law is often reactive and unable to keep pace with rapid technological developments. By incorporating the values of the *maqasid* into the legal structure, regulations can be directed not only to act but also to provide a legal framework that is preventative, adaptive, and educational. For example, this can be done through increasing understanding of value-based digital law, increasing the responsibility of platform providers, and establishing ethical and community-based digital content monitoring institutions. This approach ultimately provides a new paradigm: law functions not only as a tool of social control but also as a moral and spiritual instrument that creates a just, humane, and sustainable digital ecosystem.

*C. Digital Literacy and Islamic Ethics Education as Preventive Measures*

Literature reveals that the majority of deepfake abuse incidents are caused by a lack of digital literacy, a weak ability to verify information, and a lack of ethical knowledge when interacting online. Therefore, the most effective prevention method relies not only on detection technology or legal action, but also on digital character education based on moral values from an early age. This finding strongly emphasizes the strategic importance of the *tabayyun* principle within the proposed Islamic ethical framework. In the context of deepfake prevention, *tabayyun* transcends mere technical fact-checking; it demands a moral responsibility from every user to be diligent in examining the of information ṣidq before sharing or acting upon it. This principle serves as the cornerstone for digital character education, encouraging the cultivation of *amanah* in the digital sphere, both as content consumers and creators. By embedding this value early on, the framework shifts the culture of digital interaction from one driven by sensationalism and quick dissemination to one governed by prudence and ethical accountability. Consequently, the reliance on external punitive measures is reduced, and prevention is internalized, transforming the individual into the primary ethical firewall against the propagation of sophisticated disinformation and the violation of *ḥifz al-'irḍ* (protection of honor).

Integrating Islamic ethical values into digital literacy offers a preventative solution, emphasizing moral formation rather than simply transferring technical knowledge. The value of trustworthiness fosters awareness that every individual has a moral responsibility for the content they produce, consume, or disseminate. The kizb serves as an ethical foundation to prevent the spread of false, misleading, or manipulative content. Meanwhile, *Tabayyun* directs individuals to critically verify information before believing or sharing it, a crucial principle in addressing the flood of deepfake content.

Educational institutions, from elementary to higher education, must incorporate a digital literacy curriculum oriented toward the Maqasid that not only provides technical skills such as identifying fake content and maintaining digital security, but also fosters moral-spiritual awareness in the use of technology. Learning models can include case studies, simulations of deepfake incidents, digital ethics training, and collaborative learning with families and communities. This strategy not only creates individuals skilled in technology but also possesses strong character, self-control, and strong ethics in digital interactions. If implemented comprehensively, digital literacy that embraces Islamic ethics can serve as an effective social bulwark to reduce the potential misuse of deepfake technology in society.

*D. AI Ethics and Governance Standards Based on Islamic Values*

A literature review shows that establishing ethical AI governance cannot be based solely on technical or secular approaches, but must also incorporate moral and spiritual values relevant to the socio-cultural context of society. In the Indonesian context, integrating the principles of *Maqasid al-Shari'ah* provides a strong ethical foundation to ensure that the development and use of AI are carried out responsibly, fairly, and beneficial to public welfare. This approach assumes that technology is not neutral; it has moral implications that require clear ethical guidelines rooted in humanitarian values.

Maqasid principles such as justice *('adl)*, *(maṣlaḥah)*, *(amanah)*, and protection of human honor and dignity *(hifz al-'ird)* serve as a moral framework that can be applied in AI governance. The value of ('adl) emphasizes that AI systems must avoid all forms of algorithmic bias and discrimination, both against individuals and specific social groups. This is crucial given that modern artificial intelligence has the capacity to influence public opinion, inform decision-making recommendations, and determine access to digital services. Therefore, the application of the value of justice demands ethical audits, fairness checks, and regulations that ensure inclusivity at every stage of AI development.

Meanwhile, the principle of maṣlaḥah emphasizes that the development of artificial intelligence must provide clear benefits to society and not cause harm in social, moral, or psychological aspects. In this view, innovation in AI should focus on improving the quality of life, the effectiveness of public services, education, health, and social development, rather than solely pursuing technological advancement or financial gain. *maṣlaḥah* also requires ethical assessments to ensure that the use of AI does not violate human rights, digital privacy, and individual dignity.

The value of *amanah* directly impacts responsibility and transparency in technology management. Developers, policymakers, and organizations using AI are expected to maintain integrity in data processing, maintain system security, and provide honest and clear information regarding the algorithm's operation. Operationally, this principle requires developers to adopt immutable watermarking mechanisms on AI-generated content to ensure transparency and authenticity of information, as a manifestation of digital honesty grounded in *amanah*. *Amanah* also demands moral responsibility for the impacts of AI use, so that any misuse, manipulation, or exploitation can be minimized through ethically grounded oversight.

Furthermore, the principle of *ḥifz al-'ird* plays a crucial role in addressing issues of AI misuse, including deepfakes, digital identity theft, and content manipulation. This principle emphasizes that human dignity must be safeguarded in every technological application. Therefore, AI ethics policies based on *maqasid* must encourage preventative measures against digital identity exploitation, the dissemination of degrading content, and privacy violations that threaten personal integrity in the digital environment. By integrating these values, the formulated AI management focuses not only on innovation, effectiveness, and technological advancement but also on sensitivity to the moral-spiritual dimensions of society. This method is not intended to replace global AI ethical norms, such as those promoted by UNESCO, the OECD, or the European AI Act, but rather to complement them. The Islamic ethical aspect adds a more humane and comprehensive level of morality, aligning with the religious character of Indonesian society, which values human values.

Overall, AI management based on *Maqasid al-Shari'ah* has the potential to build a technology ecosystem that is not only safe and fair, but also morally sustainable. This method provides an opportunity for Indonesia to create a unique, contextual AI ethical model that can serve as an international reference as an exemplary application of technological ethics based on spiritual values, humanity, and social justice.

### E. The Social Impact of Deepfakes and Community Resilience

The results of the literature review indicate that the misuse of deepfake technology has broad and complex social consequences for societal structures, encompassing psychological, social, legal, and moral elements. Deepfakes not only impact individual victims but also threaten the moral health of the community as a whole. Victims of deepfakes, particularly those affected by pornographic deepfakes, experience serious reputational damage, mental distress, emotional trauma, and even social exclusion due to defamation and inaccurate digital identity changes. In a social context still imbued with patriarchal culture, women who are victims often experience greater stigma, demonstrating that the effects of deepfakes are not gender-neutral.

Beyond causing harm to individuals, this phenomenon also erodes public trust in the digital space. The public finds it increasingly difficult to distinguish between true and false information, which can lead to a crisis of trust that undermines social harmony. This decline in trust in the media, public institutions, and authority figures can potentially lead to polarization, radicalization of views, and a growing culture of mutual suspicion. In the long term, the spread of manipulated content can lead to moral confusion, where lies become commonplace and ethical boundaries become increasingly blurred.

In response to this challenge, strengthening digital resilience should be a primary strategy to minimize the social impact of deepfakes. An Islamic approach emphasizes that resilience is not only cognitive but also encompasses spiritual, moral, and social aspects. Internalizing Islamic values such as *(tabayyun)*, *(amanah)*, and preventing evil in the digital world *(nahi munkar)* can improve society's ability to recognize manipulation and avoid the spread of harmful content.

Strengthening resilience must be carried out collaboratively and at various levels. Within the family context, teaching digital ethics from an early age is crucial. Institutionally, schools and universities must incorporate digital literacy related to ethics into their learning programs. Meanwhile, community organizations, religious leaders, and Islamic outreach institutions play a crucial role as guardians of societal digital norms by providing moral guidance, spaces for discussion, and support for communities. This approach ensures that handling the deepfake issue is not solely punitive through regulations, but also preventive through character development and the development of the community's spiritual aspects.

### F. Religious Policy Reform in Response to Deepfake Technology

The study's findings also indicate that religious institutions play a crucial role in addressing today's technological advances, including deepfakes and artificial intelligence. However, this involvement requires changes in religious policies to prevent them from being reactive, normative, and slow to keep pace with technological innovation. Religious institutions need to conduct modern such as Islamic ijtihad *(ijtihad)* so that Islamic values remain applicable, relevant, and provide solutions in confronting the increasingly advanced digital world. The necessity for contemporary *ijtihād* (independent reasoning or interpretation) arises because the harms associated with deepfakes—such as systemic disinformation, psychological trauma, and reputational collapse—are nawāzil (novel issues) that were unknown to classical jurisprudence. Traditional legal rulings often focus on material proof, which is difficult to establish in the fluid, non-physical realm of deepfakes. Therefore, religious institutions must mobilize scholars to formulate fatwā (legal opinions) and normative guidelines that directly address the digital manifestation of moral transgressions like kidhb (falsehood) and ghībah (slander). This proactive, modernized *ijtihād* ensures that the preventive objectives of *Maqāṣid al-Sharī'ah* remain the core of digital ethics. By issuing clear and timely religious policy that integrates the principles of *ḥifz al-'ird* and *tabayyun*, religious bodies can effectively guide the public and technology practitioners, transforming the Islamic ethical framework from an academic concept into a dynamic, practical tool for AI governance.

Religious policy transformation needs to be carried out through the development of ethical guidelines, fatwas, and normative frameworks that clarify moral boundaries for society in using digital technology. Prohibiting the spread of digital slander *Fatwas* , distorted content, and privacy must be expanded and updated to encompass current issues such as digital identity manipulation, image authenticity, and the right to dignity in the virtual space. *Fatwas* should not only focus on what is permissible *(halal)* and what is forbidden *(haram)* but also include practical guidance that can be implemented by the public, especially the younger generation and digital content activists.

Religious policies should strengthen the position of Muslim scholars and academics in discussions on technology through cross-disciplinary collaboration. The integration of sharia science, technology, and ethics is crucial for the resulting policies to have theological depth and technical relevance. The involvement of AI practitioners, digital forensics experts, psychologists, and sociologists is essential to create comprehensive religious guidance that is responsive to digital dynamics. The Indonesian Ulema Council (MUI), Islamic mass organizations, and religious educational institutions can play a key role in this transformation.

More deeply, the overhaul of religious policy is not only intended to provide ethical guidance but also to encourage fundamental change in the digital environment. Religious institutions can play a role in advocating for public policy, conducting educational campaigns on digital ethics, and formulating *da'wah* modules that focus on current issues such as digital identity, data protection, and ethics in social media. Therefore, a faith-based approach is not limited to ethical calls alone, but serves as a driver of social change that builds an ethical, moral, and civilized digital culture.

In general, religious policy reform plays a role in strengthening Islam's position as a relevant moral reference in the age of AI and deepfakes. This method ensures that the digitalization process does not distance society from the principles of faith, but rather becomes a tool to strengthen the spirituality, humanity, and ethical responsibility of the community in facing today's technological challenges.

### G. Challenges of Implementing Islamic Ethics

Although the Islamic ethical framework based on *Maqasid al-Shari'ah* offers holistic preventive solutions, its implementation in the real world is not without challenges. Researchers identify several potential key obstacles:

1) *Jurisdictional Challenges and Global Adoption:* The first challenge is jurisdictional. Encouraging multinational technology companies (Big Tech) operating within a secular legal framework to adopt ethical auditing based on *Maqasid al-Shari'ah* requires significant diplomatic and advocacy efforts. Without regulatory incentives or market pressure, adoption of these principles will likely be voluntary and limited.
2) *Legal Harmonization:* There is a potential conflict between interpretations of Islamic principles, such as *hifz al-'ird*, and universal human rights legal frameworks or secular national laws on freedom of expression. Finding an appropriate balance to ensure that protection against defamatory deepfakes does not evolve into a repressive form of censorship remains a complex legal challenge.
3) *Technological Arms Race:* In the technology domain, solutions like *watermarking* (as a form of *trust*) face the challenge of a constant *adversarial arms race*. As rapidly as detection and *watermarking* technologies advance, so does the *deepfake* technology designed to evade such detection. Technical implementation requires ongoing industry collaboration and mutually agreed-upon standards.

Addressing these challenges requires a multi-stakeholder approach involving not only scholars and engineers, but also policymakers, non-governmental organizations, and global business entities.

The findings of the synthesis indicate that the integration of Islamic ethical values, particularly through the *Maqasid al-Shari'ah*, provides a far more comprehensive preventive framework for addressing the misuse of deepfake technology than currently common AI ethics approaches. Western approaches tend to emphasize justice, transparency, and responsibility, but are reactive, technocratic, and remain concentrated on technological aspects and formal regulations. Their philosophical weakness lies in their strong focus on (ex-post) detection and formal sanctions, making them less effective in addressing the issues of intent and moral motivation behind the creation of harmful deepfakes such as *kizb*, and *tadlis*. In contrast, Islamic ethical principles operate at the level of moral character building, spiritual awareness, and the creation of a digital culture of integrity dimensions that have so far received less attention in discussions of AI ethics globally. The Islamic approach offers a holistic solution by addressing the dimensions of *maṣlaḥah* and harm *mafsadah* as primary moral considerations.

One significant contribution of the literature is its emphasis that the misuse of deepfake technology is not merely a technological or legal issue, but rather a moral and social issue that directly impacts individual dignity and public trust. The *Maqasid al-Shari'ah* framework provides a values-based approach encompassing the protection of life, intellect, wealth, lineage, and honor. In the context of deepfakes, the dimension of protection of honor and reputation, grounded in the principle of *hifz al-'ird* becomes particularly relevant, as this technology is often misused to damage a person's dignity, reputation, and privacy, even without physical involvement. This perspective deepens the perspective on AI ethics, as it establishes that digital violations of visual identity also constitute a violation of human dignity.

The discussion analysis also shows that handling deepfakes cannot rely solely on regulations and technology to detect them. Digital literacy based on ethical values is an important aspect increasingly proposed by researchers. Related research emphasizes that digital character education, understanding the values of trust, integrity, and the culture of religious reflection grounded in *tabayyun* can increase social resilience against the manipulation of counterfeit content. Therefore, strengthening ethical awareness from an early age and within an educational context can produce a generation that is more sensitive, critical, and possesses moral integrity in the use and dissemination of digital information.

In addition to moral education, governance and public policy aspects are also of strategic importance. The literature shows that the Indonesian legal system still has gaps in addressing deepfake cases, particularly regarding non-material losses such as defamation and psychological impacts. This is where the principles for the Law of God *Maqasid* and Islamic ethics can serve as a normative basis for developing more humane and comprehensive policies. This approach focuses not only on punishment for violators but also on restoring the dignity of victims, protecting privacy, and preventing similar incidents from recurring. Furthermore, the discussion suggests that strengthening the ethical ecosystem in the field of AI must be driven by collaboration between academics, government, educational institutions, religious leaders, and digital platforms. The creation of national ethical standards that combine universal principles and local values, including Islamic values accepted by the majority of Indonesian society, is believed to increase the effectiveness of existing policies. This is also in line with global trends recognizing that AI ethics cannot be considered universal and must consider cultural context, societal values, and local beliefs.

From a social perspective, this paper also emphasizes the damaging effects of deepfakes on public trust and social relations. The crisis of trust created by deepfakes can undermine trust in societal interactions, trigger conflict, and create social instability. Therefore, building a digital culture based on trust, responsibility, and prudence is not merely normative but a necessity for maintaining social cohesion amidst the development of AI. Overall, this discussion emphasizes that the Islamic ethical approach does not seek to replace the Western ethical framework, but rather complements it with spiritual, moral, and cultural aspects that have received less attention. Synergy between the two can create a more comprehensive ethical model: the Western approach is able to provide technical mechanisms and standards that.

## V. Conclusion

The development of AI technology, particularly deepfakes using GANs, has caused significant transformations in the digital realm worldwide. AI's ability to create highly similar audio-visual content not only supports advancements in the creative industries, learning, reconstruction of past events, and entertainment technology, but also poses complex new risks to the veracity of information and public beliefs. In a social context, the proliferation of deepfakes has exacerbated epistemic problems, blurring the distinction between truth and fabrication, impacting public perception, decision-making processes, and social balance. In a personal context, this technology creates opportunities for digital identity abuse, including unauthorized use of a person's face or voice for pornography, threats, financial fraud, and personal reputational damage. The complexity of these risks indicates that the problem of deepfakes is not merely technical but also involves much deeper moral, legal, and social dimensions.

This research concludes that the current ethical foundations of AI, largely based on secular norms and technical methods such as detection and validation, are limited and tend to be applied after the fact. This foundation fails to address the core of the problem, which includes the perpetrator's intent, the user's moral integrity, and the values that can prevent abuse before it occurs. Therefore, there is an urgency for a new ethical model that not only addresses consequences but also shapes actions, moral awareness, and social responsibility from the outset. This literature review demonstrates that Islamic ethics offers a more comprehensive, preventative model, and is appropriate for today's digital era through the principles of *Maqāṣid al-Sharī'ah*.

This study found that the misuse of deepfakes substantially contradicts the six main pillars of *Maqāṣid al-Sharī'ah*: Protection of religion *(ḥifẓ al-dīn)*, *(ḥifẓ al-nafs)*, Reason *(ḥifẓ al-'aql)*, *(ḥifẓ al-nasl)*, Property *(ḥifẓ al-māl)*, and especially honor and dignity *ḥifẓ al-'irḍ*. The results reinforce the importance of incorporating Islamic values into technology governance to provide a moral perspective not found within conventional ethical frameworks.

This study identified three pillars of strategic intervention based on Islamic ethical principles:

1) Regulation and Law: Current regulations, including provisions in the Information and Electronic Transactions Law, do not explicitly protect honor and focus more on material losses. To allow for recognition of immaterial losses, including reputational damage, psychological trauma, and the social impact experienced by victims, this study recommends legislative changes. Furthermore, laws based on *Maqasid* encourage the state to not only punish perpetrators but also comprehensively protect victims.

2) Technology Governance: To ensure that the development of artificial intelligence is in accordance with human moral values, technological ethics must be based on the values of *adl*, *amanah*, and *ṣidq*. Conducting ethical audits of AI is crucial, especially for high-risk generative models. The application of Islamic ethics in technology governance can foster transparency, accountability, and fairness, thereby preventing the misuse of deepfakes through measures such as irreversible watermarking, dataset transparency, and algorithmic fairness.

3) Digital Literacy and Education: Prevention efforts will not be successful without a society that can exercise restraint, verify, and clarify information before spreading it. The principle of tabayyun, verification, and checking the truth, is highly relevant for digital literacy in the era of visual manipulation. Islamic values-based education has the potential to raise public moral awareness and serve as a social bulwark against the misuse of technology.

Overall, this research makes a significant contribution by providing a functional and normative ethical framework. It shifts the paradigm for handling deepfakes from a sanction-and-detection-based approach to a preventative approach centered on moral and spiritual values. This framework encourages technological advancement that does not violate humanity, justice, and the maslahah and strengthens society's digital resilience. This aligns with the Islamic perspective on technology as a tool for good rather than harm.

However, this research is merely conceptual and does not empirically test the application of this ethical framework in the real world. Therefore, further research is strongly recommended. Examples include a case study on the application of *Maqasid* in AI model training, the design of an ethical audit checklist based on Islamic principles, a survey based on the principle of *tabayyun* on the level of public acceptance of digital literacy, or the development of an ethical framework based on Islamic principles. These empirical methods would enhance the applicability and validity of the framework.

Overall, this research demonstrates that Islamic ethics is not only relevant but also essential for addressing the moral issues raised by deepfake technology. There is a strong possibility that these values will become a new foundation for building fair, sustainable, and human-first AI governance in the digital age.

ACKNOWLEDGMENT

Although this research has only been partially completed during this period, the author would like to express sincere gratitude to the Department of Informatics, Universitas Islam Negeri (UIN) Sunan Gunung Djati Bandung, as the supporting institution of this research. The academic guidance, institutional support, and access to adequate facilities provided by the university significantly contributed to the smooth progression of the research process. Furthermore, the author appreciates the constructive and collaborative academic environment fostered by the institution, which enabled systematic analysis, the development of more focused research ideas, and deeper engagement with the research topic. This acknowledgment is presented as an expression of appreciation for the guidance, support, and academic opportunities provided during the research preparation process. This research received no external funding.